\documentclass[twocolumn,aps,longbibliography,floatfix,showpacs,superscriptaddress]{revtex4}
\usepackage{mathrsfs,bm,amsmath}
\usepackage{graphicx}

\def\e{{\epsilon}}
\def\k{{ {\bm k} }}

\def\q{{ {\bm q} }}
\def\Q{{ {\bm Q} }}
\def\0{{ {\bm 0} }}

\def\a{{\alpha}}
\def\b{{\beta}}

\allowdisplaybreaks[4]

\makeindex

\begin{document}

\title{
Multistage Electronic Nematic Transitions 
in Cuprate Superconductors:
Functional-Renormalization-Group Analysis
}

\author{Masahisa Tsuchiizu}
\affiliation{Department of Physics, Nara Women's University, 
Nara 630-8506, Japan}
\affiliation{Department of Physics, Nagoya University, 
Nagoya 464-8602, Japan}

\author{Kouki Kawaguchi}
\author{Youichi Yamakawa}
\author{Hiroshi Kontani}
\affiliation{Department of Physics, Nagoya University, 
Nagoya 464-8602, Japan}

\date{\today}

\begin{abstract}
Recently, complex rotational symmetry breaking phenomena
have been discovered experimentally
in cuprate superconductors.
To find the realized order parameters, we study various unconventional
charge susceptibilities in an unbiased way, 
by applying the functional-renormalization-group method to the 
$d$-$p$ Hubbard model.
Without assuming the wavevector of the order parameter, we reveal that the
most dominant instability is the uniform ($\bm q=\bm 0$) charge
modulation on the $p_x$ and $p_y$ orbitals, which possesses the $d$-symmetry.
This uniform nematic order triggers another nematic 
$p$-orbital density wave along the axial (Cu-Cu) direction
at $\Q_{\rm a}\approx(\pi/2,0)$.
It is predicted that uniform nematic order 
is driven by the spin fluctuations in the pseudogap region, 
and another nematic density-wave order at $\q=\Q_{\rm a}$
is triggered by the uniform order.
The predicted multistage nematic transitions
are caused by the Aslamazov-Larkin-type fluctuation-exchange processes.
\end{abstract}


\maketitle

\section{Introduction}

\begin{figure}[b]
\includegraphics[width=8cm]{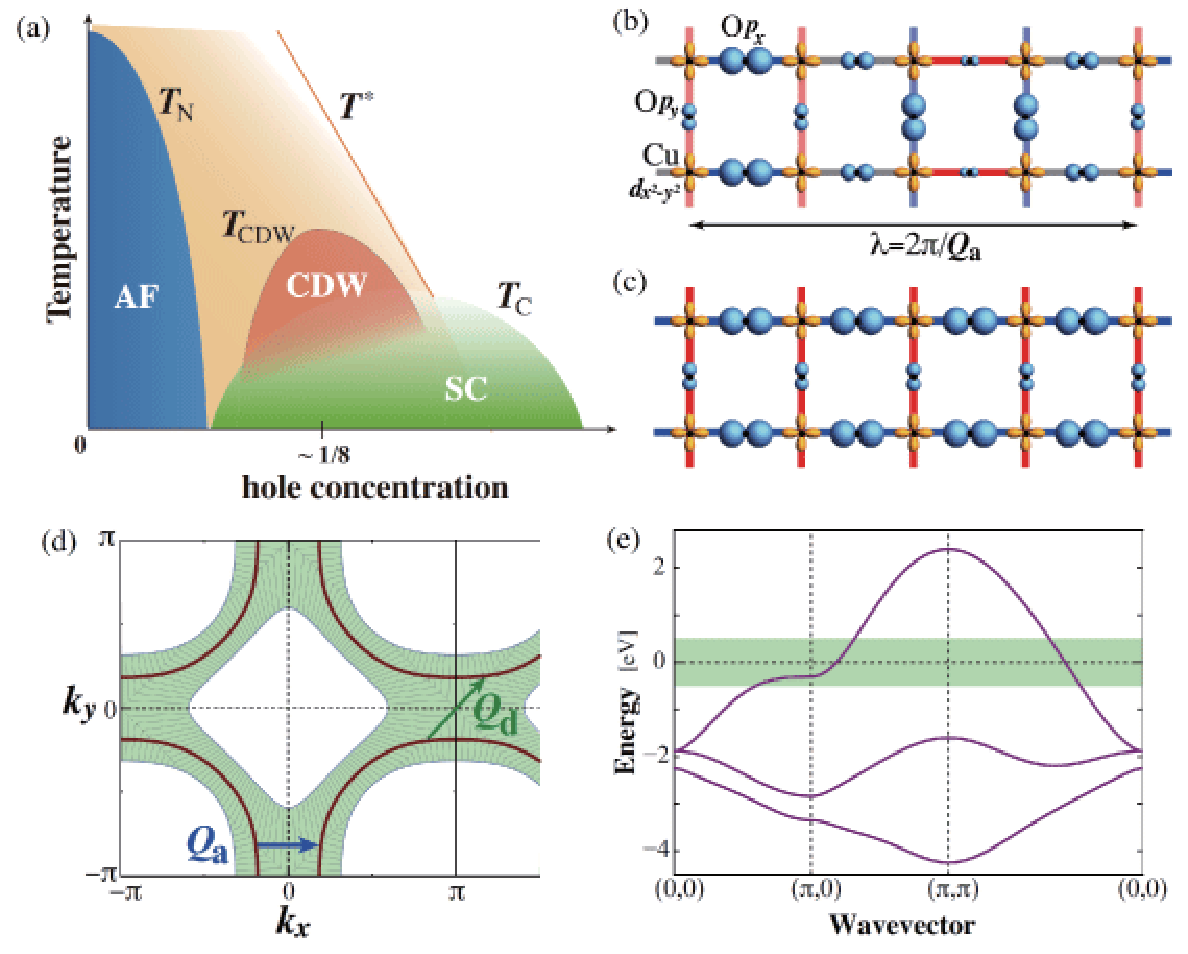}

\caption{
(a) Schematic phase diagram of the high-$T_c$ cuprate superconductors.
$T^*$, $T_{\rm CDW}$, $T_N$, and $T_{\rm c}$ are the transition temperatures
for the pseudogap state, CDW order, magnetic order, and superconductivity, 
respectively.
We study 10\% doping case shown by the vertical broken line.
(b) Schematic charge distribution in the 
$d$-symmetry $p$O-CDW state with
 the wavevector $\bm q= \bm Q_\mathrm{a} \approx (0.5\pi, 0)$.
(c) The uniform nematic $p$O-CDW state with $n_x\ne n_y$.
(d) The FS and 
(e) the energy dispersion of the present $d$-$p$ model.
The lower energy region ($|E|<\Lambda_0=0.5$ eV) is divided into
the $N_p=128$ patches to perform the RG analysis.
}
\label{fig1}
\end{figure}

In the normal state of high-$T_c$ cuprate superconductors,
interesting unconventional order parameters emerge
due to the strong interference among the spin, charge, and orbital degrees of freedom.
These phenomena should be directly related to the 
fundamental electronic states in the pseudogap region.
The emergence of the charge-density-wave (CDW) states
inside the pseudogap region has been confirmed by 
the x-ray and STM measurements
\cite{
Ghiringhelli2012sc%
,Chang:2012ib%
,Fujita:2014kg%
,Wu:2015bt%
,Hamidian:2015eo%
,Comin:2016ho%
}, 
as schematically shown in Fig.\ \ref{fig1}(a).
The observed CDW pattern is shown in Fig.\ \ref{fig1}(b),
in which the density modulations mainly occur on the oxygen $p_x$ and $p_y$
orbitals with antiphase ($d$-symmetry) form factor.
The discovery of the CDW has promoted
significant progress in the theoretical studies,
such as the spin-fluctuation-driven density-wave scenarios
\cite{Davis:2013ce,Metlitski:2010gf,Husemann:2012eb,Efetov:2013ib,Sachdev:2013bo,Mishra:2015fb,Yamakawa:2015hb,Orth:2017}
and the superconducting-fluctuation scenarios
\cite{Berg:2009gt,Fradkin:2015co,Wang:2015iq,Lee:2014ka}.

The origin and nature of the pseudogap phase below $T^*$ remain unsolved.
For example, it is unclear 
whether the pseudogap is a distinct phase or a continuous crossover.
The short-range spin-fluctuations at $T\sim T^*$
induce the large quasiparticle damping \cite{TPSC,Kotliar,Maier},
which causes the pseudogap in the density-of-states.
On the other hand,
the phase transition around $T^*$
have been reported by the resonant ultrasound spectroscopy 
\cite{Shekhter:2013eh},
ARPES analysis \cite{ARPES-Science2011}, and 
magnetic torque measurement \cite{Matsuda:2016}.
Especially, Ref. \cite{Matsuda:2016} discovered the 
$C_4$ symmetry breaking (nematic) transition, and
its natural candidate is the uniform CDW with $d$-symmetry
schematically shown in Fig.\ \ref{fig1}(c).
Then, a fundamental question is
what mechanism can account for such unconventional 
multistage CDW transitions.
No CDW instabilities are given by the 
mean-field-level approximations, like the 
random-phase-approximation (RPA),
unless large inter-site interactions are introduced
\cite{Bulut,Yamakawa:2015hb}.
Therefore, higher-order many body effects, called the vertex corrections (VCs),
should be essential for the CDW formation
\cite{
Onari:2012jb,Davis:2013ce,Metlitski:2010gf,Husemann:2012eb,
Efetov:2013ib,Sachdev:2013bo,Mishra:2015fb,Yamakawa:2015hb}.

In many spin-fluctuation-driven CDW scenarios,
the CDW wavevector is given by the 
minor nesting vector $\bm Q_\mathrm{a}$ or $\bm Q_\mathrm{d}$
in Fig.\ \ref{fig1}(d);
$\bm Q_\mathrm{a}$ is the ``axial-wavevector'' parallel to 
the nearest Cu-Cu direction, and
$\bm Q_\mathrm{d}$ is the ``diagonal-wavevector''.
The experimental axial CDW is obtained 
if the Aslamazov-Larkin VCs (AL-VCs) are taken into account
\cite{Yamakawa:2015hb}.
In addition,
the uniform ($\q={\bm 0}$) CDW instability
has been studied intensively based on the Hubbard models
\cite{Halboth:2000vm%
,Honerkamp:2005fv%
,Husemann:2012eb
,Su:2011co
,Kitatani:2016ur
}.
In these studies, however, it was difficult to exclude
the possibility that the CDW susceptibility has the maximum at
finite $\q$.

Theoretically, it is difficult to 
analyze the spin and charge susceptibilities with general wavevector
$\bm q$ on equal footing, by including the VCs in an unbiased way.
For this purpose, in principle, 
the functional renormalization-group (fRG) method would be
the best theoretical method.
%
The pioneering fRG studies
\cite{%
Honerkamp:2005fv%
,Husemann:2012eb
}
were performed only in the weak-coupling region,
so the obtained CDW instability is small and 
its $\q$-dependence is not clear.
In order to overcome this problem, 
we have to improve the numerical accuracy of the fRG method,
and apply it to the two-dimensional Hubbard model 
in the strong-coupling region.

In this paper,
we study the orbital-dependent spin and charge susceptibilities 
for various symmetries on equal footing,
by analyzing the higher-order VCs in an unbiased way
using the improved fRG method.
We find that the uniform CDW 
accompanied by the $p$-orbital polarization ($n_{x}\ne n_{y}$),
shown in Fig. \ref{fig1}(c),
is driven by the antiferro spin fluctuations.
In this uniform nematic CDW phase,
another nematic CDW instability emerges at the wavevector $\q=\Q_{\rm a}$
as shown in Fig. \ref{fig1}(b).
The present study indicates that the 
uniform $p$-orbital polarization appears 
in the pseudogap region,
and the axial $\q=\Q_{\rm a}$ CDW is induced at $T_{\rm CDW}<T^*$.
These multistage CDW transitions in under-doped cuprates
originate from the higher-order AL-type VCs.

In the present study, we use the 
functional RG + constrained RPA (RG+cRPA) method.
The advantage of this method had been explained in Refs.
\cite{Tsuchiizu:2013gu,Tsuchiizu:2015cs,Tsuchiizu:2016ix,Tazai:2016vd}
and Appendix A in detail.

\section{Model and Theoretical Method}

Here, we study a standard three-orbital $d$-$p$ Hubbard model
\cite{Yamakawa:2015hb,Tsuchiizu:2016ix,Hansmann:2014ib}
expressed as
$H=\sum_{\bm k, \sigma} \bm c_{\bm k, \sigma}^\dagger \,
\hat h_0(\bm k)  \, \bm c_{\bm k, \sigma}^{}
+U\sum_{\bm j} n_{d, \bm j,\uparrow} n_{d, \bm j,\downarrow}$, 
where $\bm c_{\bm k, \sigma}^\dagger=
(d_{\bm k,\sigma}^\dagger, p_{x,\bm k,\sigma}^\dagger, 
 p_{y,\bm k,\sigma}^\dagger)$
is the creation operator for the electron 
on $d$, $p_x$, and $p_y$ orbitals,
and $\hat h_0(\bm k)$ is the kinetic term
given as the 0MTO model in Refs. \cite{Hansmann:2014ib,note}.
(The numerical results are unchanged if 
another realistic 1MTO model is used; see Appendix B.)
$U$ is the Hubbard-type on-site Coulomb interaction for the $d$ orbital,
and $n_{d,\bm j,\sigma}=d^\dagger_{\bm j,\sigma}d_{\bm j,\sigma}$ at site $\bm j$.
Hereafter, we study the $10 \%$ hole doping case.
The Fermi surface (FS) and the band structure are shown in 
Figs.\ \ref{fig1}(d) and (e), respectively.

By using the RG+cRPA theory in Ref. \cite{Tsuchiizu:2016ix},
we find that the spin susceptibility for $d$-electrons,
\begin{equation}
\chi^{\mathrm{spin}}(\bm q)
=
\frac{1}{2}
\int_0^{1/T} d\tau \,
\left\langle 
S_d(\bm q,\tau)
S_d(-\bm q,0)
\right\rangle,
\end{equation}
and the $B_{1g}$-symmetry ($d$-symmetry) charge susceptibility 
for $p$-electrons,
\begin{equation}
\chi^{p\mbox{-}\mathrm{orb}}_{d}(\bm q)
=
\frac{1}{2}
\int_0^{1/T} \! d\tau \,
\left\langle 
n^{p\mbox{-}\mathrm{orb}}_d(\bm q,\tau)
n^{p\mbox{-}\mathrm{orb}}_d(-\bm q,0)
\right\rangle , 
\end{equation}
are the most enhanced susceptibilities 
\cite{Tsuchiizu:2016ix}.
Here, 
$S_d(\bm q,\tau)$ is the $d$-electron spin operator, and
$n^{p\mbox{-}\mathrm{orb}}_d(\bm{q}) \equiv  n_{x}(\bm{q}) - n_{y}(\bm{q})$
($n_{x(y)} (\bm q ) =  \sum_{\bm k, \sigma}
 p_{x(y),\bm{k},\sigma}^\dagger p_{x(y),\bm{k+q},\sigma}$) is the 
$p$-orbital charge-density-wave ($p$O-CDW) operator with $B_{1g}$ symmetry.
If $\chi^{p\mbox{-}\mathrm{orb}}_d(\bm q)$ diverges at $\bm q=\bm Q_{\rm a}$
[$\bm q=\bm 0$], the $p$O-CDW order shown in Fig.\ \ref{fig1}(b) 
[Fig.\ \ref{fig1}(c)], which is the CDW on $p$-orbitals, is realized.
We verified that the charge susceptibilities with non-$B_{1g}$-symmetries,
such as the $A_{1g}$-symmetry total charge susceptibility
for $n\equiv n_d+ n_{x}+n_{y}$, remains small even in the 
strong-coupling region  
\cite{Tsuchiizu:2016ix}.

In the RG+cRPA method,
we calculate the scattering processes involving higher-energy states
$|E_{\bm k,\nu}|>\Lambda_0$ [$\nu$ being the band index; 
see Fig.\ \ref{fig1}(e)]
using the RPA with the energy-constraint, 
and incorporate their contributions into 
the initial vertex functions of the RG equations
\cite{Tsuchiizu:2013gu,Tsuchiizu:2015cs,Tsuchiizu:2016ix,Tazai:2016vd}.
Using the RPA, the higher-energy processes
are calculated accurately by dropping the VCs,
which are less important for higher-energy processes.
The lower-energy scattering processes for $|E_{\k,\nu}| < \Lambda_0$
are calculated by solving the RG equations,
based on the $N_p$-patch RG scheme
\cite{Halboth:2000vm,Metzner:2012jv}.
Hereafter, we put $N_p=128$ and $\Lambda_0=0.5$ eV.
In the RG+cRPA method, the numerical accuracy of the susceptibilities
is greatly improved even in the weak-coupling region
since the cRPA is used for the higher-energy processes,
for which the $N_p$-patch RG scheme is less accurate.
We verified that the numerical results are essentially 
independent of the choice of $\Lambda_0$
when  $E_{\rm F}\gtrsim \Lambda_0\gg T$.

By solving the RG equations,
many-body vertices are gradually renormalized as
reducing the energy scale $\Lambda_l=\Lambda_0e^{-l}$ 
with increasing $l\ (\ge0)$.
In principle, the renormalization of the vertex saturates when
$\Lambda_l$ reaches $\sim T$
\cite{Metzner:2012jv,Bourbonnais:2004review}.
Here, we introduce the lower-energy cutoff $\Lambda_{\rm low}\ (\sim T)$
in the RG equations for the four-point vertex $\Gamma_l^{s(c)}$,
and stop the renormalization at $\Lambda_l=\Lambda_{\rm low}$;
see Appendix A and Ref. \cite{Husemann:2012eb}.
(We do not introduce the lower-energy cutoff 
in the RG equations for $\chi^{s,c}(\q)$.)
In the previous study \cite{Tsuchiizu:2016ix}, 
we set a large cutoff $\Lambda_{\rm low}=\pi T$
to achieve stable numerical results.
When $\Lambda_{\rm low}\gg T$,
the uniform ($\q={\bm 0}$) nematic susceptibility is 
especially underestimated compared to $\q\ne{\bm0}$ instabilities,
as we will discuss later.
Since we have improved the numerical accuracy 
in solving the RG equations, we can use a smaller
natural cutoff $\Lambda_{\rm low}=T$.
For this reason, we can obtain the $\q$-dependence
of the susceptibility accurately, including $\q\approx{\bm 0}$.

We find that the numerical accuracy and stability
are improved by employing the Wick-ordered scheme of the 
fRG formalism, in which the 
cutoff function $\Theta_<^\Lambda(\e)= \Theta(\Lambda-|\e|)$
is used for the Green function \cite{Metzner:2012jv}.
In this scheme, in principle, the VCs due to the higher-energy processes 
are included more accurately,
compared to using another cutoff function 
$\Theta_>^\Lambda(\e)= \Theta(|\e|-\Lambda)$
based on the Kadanoff-Wilson scheme 
used in Ref. \cite{Tsuchiizu:2016ix}.

\begin{figure}[t]
\includegraphics[width=8cm]{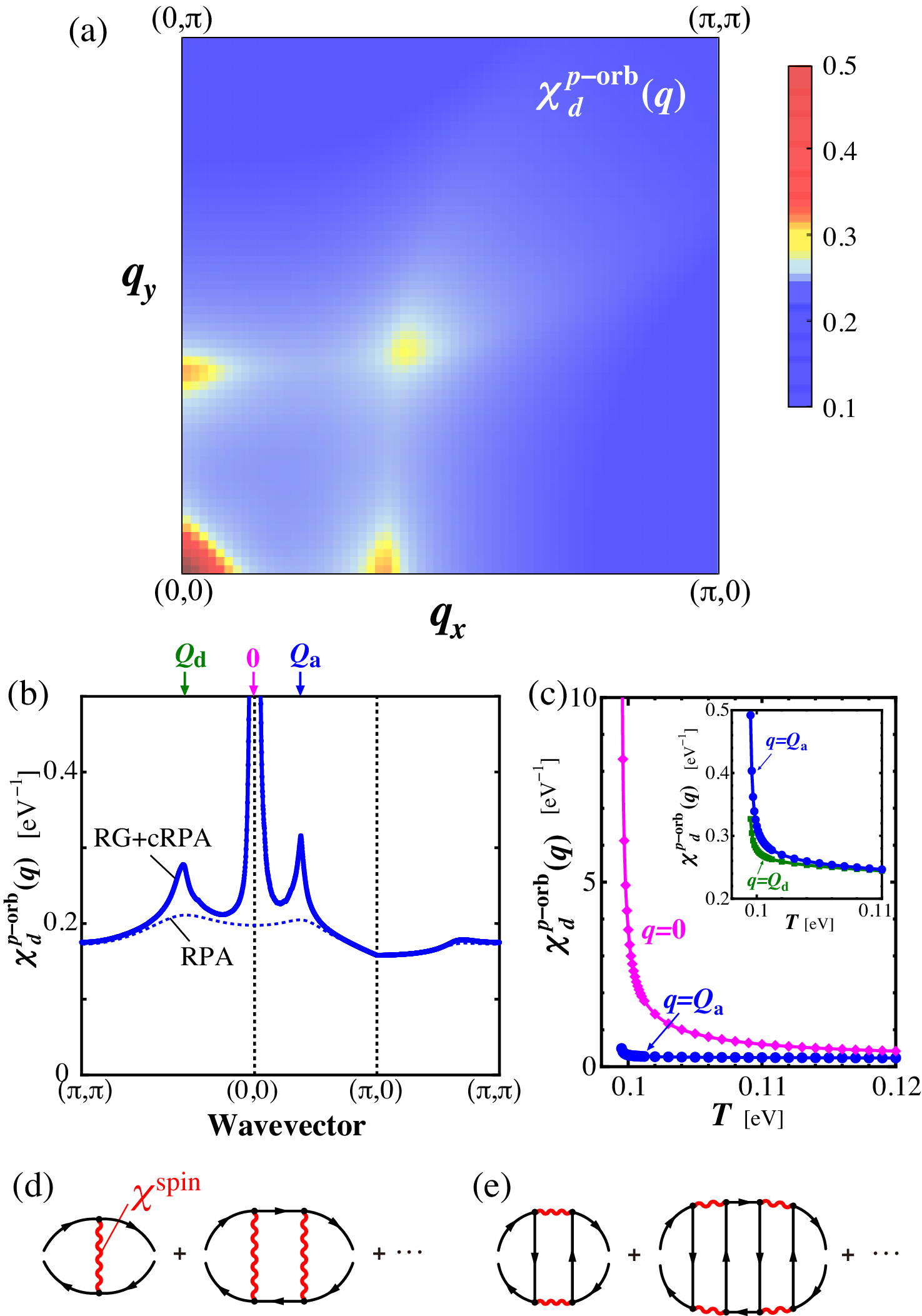}

\caption{
(a) (b) The RG+cRPA result of the $p$O-CDW susceptibility 
$\chi^{p\mbox{-}\mathrm{orb}}_d(\bm q)$ obtained for $U=4.32$ eV at $T=0.1$ eV.  
The RPA result is also shown for comparison in (b).
The axial wavevector is $\bm Q_\mathrm{a}\approx  (0.37\pi, 0)$ 
and the diagonal wavevector is
$\bm Q_\mathrm{d}\approx  (0.40\pi,0.40\pi)$.  
Both $\bm Q_\mathrm{a}$ and $\bm Q_\mathrm{d}$ 
correspond to the wavevector connecting the hot spots shown 
in Fig.\ \ref{fig1}(b).  
(c) The $T$-dependence of $\chi^{p\mbox{-}\mathrm{orb}}_d(\bm q)$ for $U=4.32$ eV.
(d) VCs due to the MT processes.
(e) VCs due to the AL processes.
}
\label{fig2}
\end{figure}

\section{Multistage Electronic Nematic Transitions}

In Figs.\ \ref{fig2}(a) and (b), we show the 
$p$O-CDW susceptibility $\chi^{p\mbox{-}\mathrm{orb}}_{d}(\bm q)$
given by the RG+cRPA method for $U=4.32$ eV at $T=0.1$ eV.  
The obtained large peaks at 
$\bm q=\bm 0$, $\bm Q_\mathrm{a}$, and $\bm Q_\mathrm{d}$
originate from the VCs, since 
the RPA result is less singular as seen in Fig.\ \ref{fig2}(b).
As shown in Figs.\ \ref{fig2}(a)-\ref{fig2}(c), the most 
dominant peak locates at $\bm q=\bm 0$.
This is consistent with the experimental uniform
nematic transition at $T^* \ (>T_{\rm CDW})$
\cite{Matsuda:2016}.
We also obtain the 
peak structures at $\bm q=\bm Q_\mathrm{a}$ and $\bm Q_\mathrm{d}$, 
consistently with our previous fRG study \cite{Tsuchiizu:2016ix}.
Figure \ref{fig2}(c) shows that
$\chi^{p\mbox{-}\mathrm{orb}}_d(\bm 0)$  monotonically increases
with decreasing $T$,
consistently with the recent electronic nematic 
susceptibility measurement \cite{Shibauchi-nem}.
At low temperatures,
$\chi^{p\mbox{-}\mathrm{orb}}_d(\bm Q_\mathrm{a})$ increases steeply
and becomes larger than $\chi^{p\mbox{-}\mathrm{orb}}_d(\bm Q_\mathrm{d})$,
shown in the inset of Fig.\ \ref{fig2}(c).
Note that the temperature $T=0.1$ eV is comparable to $T^*\sim300$ K
if the mass-enhancement factor $m^*/m_{\rm band}\sim 3$ is taken into account.

\begin{figure}[t]
\includegraphics[height=8.cm]{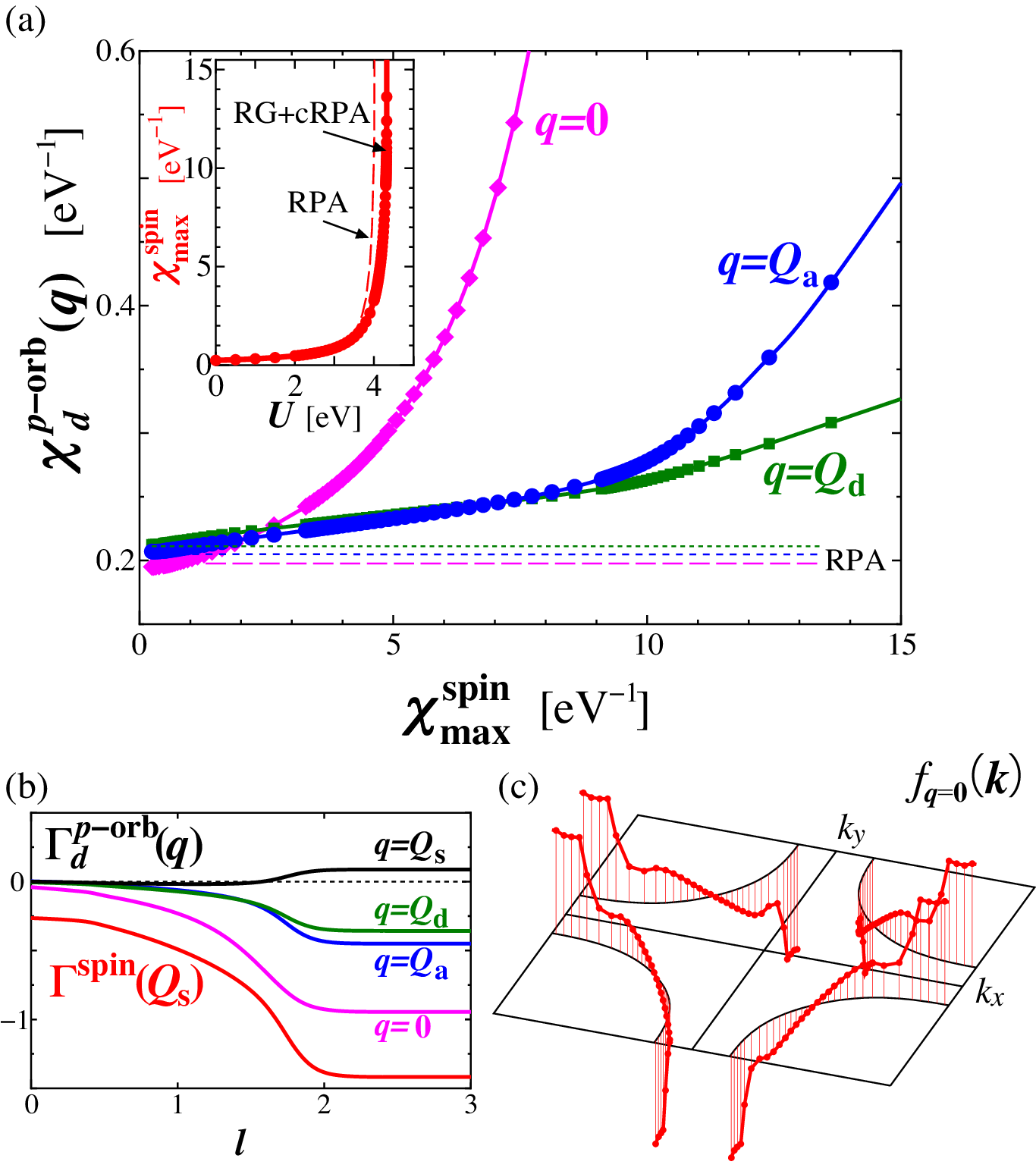}
\caption{
(a) 
The RG+cRPA result of $\chi^{p\mbox{-}\mathrm{orb}}_d(\bm q)$
at three peak positions
as a function of $\chi^\mathrm{spin}_\mathrm{max}$ ($\Delta E_{p}=0$).
The RPA results are also shown by lines.
In the inset, the $U$ dependence of 
$\chi^\mathrm{spin}_\mathrm{max}$ is shown.
(b)
Scaling flows of the effective four-point vertices 
for the $p$O-CDW with $d$ symmetry, for $U=4.32$ eV at $T=0.1$ eV. 
$l\ (\ge0)$ is the scaling parameter.
The scaling flows for spin channel is also shown where
$\Q_\mathrm{s}$ is the nesting vector
$\approx (\pi, \, 0.78 \pi)$ or $(0.78\pi, \, \pi)$.
(c) The optimized form factor $f_{\q=\bm0}(\k)$ on the FS, 
which has the $d$-symmetry.
}
\label{fig3}
\end{figure}

The enhancement of $\chi^{p\mbox{-}\mathrm{orb}}_d(\bm q)$
is caused by the spin fluctuations,
due to the strong charge-spin interplay given by the VCs.
The moderate peak at $\bm Q_\mathrm{d}$ is caused by the 
Maki-Thompson (MT)-VCs, given by the series of the
single-fluctuation-exchange processes shown in Fig. \ref{fig2}(d)
 \cite{Sachdev:2013bo,Mishra:2015fb}.
However, the MT-VCs cannot account for the dominant 
peaks at $\bm q=\bm 0$ and $\bm Q_\mathrm{a}$.
Recently,
it was found that the uniform nematic order
in the Fe-based superconductors
\cite{Onari:2012jb,Kontani:2014ws} and Sr$_3$Ru$_2$O$_7$
\cite{Ohno:2013hc,Tsuchiizu:2013gu}
is driven by the AL-VC,
given by the series of the double-fluctuation-exchange processes
shown in Fig. \ref{fig2}(e).
In fact, the first term in Fig. \ref{fig2}(e)
is proportional to $\sum_\k\chi^{\rm spin}(\k+\q)\chi^{\rm spin}(\k)$,
which takes large value for $\q={\bm 0}$ when 
$\chi^{\rm spin}_{\rm max}\gg1$
 \cite{Onari:2012jb,Grilli}.
Later, we will demonstrate that the AL-VC
causes the uniform and axial CDW instabilities in the present $d$-$p$ model.

Next, we investigate the $U$-dependences of the susceptibilities.
In the inset of Fig.\ \ref{fig3}(a), we show the $U$ dependence of 
$\chi^\mathrm{spin}_{\rm max}\equiv {\rm max}_\q\{\chi^\mathrm{spin}(\q)\}$.
Thanks to the numerical accuracy of the RG+cRPA method,
$\chi^\mathrm{spin}_\mathrm{max}$ perfectly follows the RPA result
for wide weak-coupling region ($U<4$ eV).
To clarify the close interplay between spin and orbital fluctuations,
we plot the peak values of 
$\chi^{p\mbox{-}\mathrm{orb}}_d(\bm q)$ as a function of 
$\chi^\mathrm{spin}_{\rm max}$ in Fig.\ \ref{fig3}(a).
In contrast to $\chi^{\rm spin}_{\rm max}$,
$\chi^{p\mbox{-}\mathrm{orb}}_{d}(\bm q)$ strongly deviates from the RPA result,
indicating the significance of the VCs.
With increasing $U$,
the peak position of $\chi^{p\mbox{-}\mathrm{orb}}_{d}(\bm q)$ 
shifts to $\q={\bm 0}$ at $\chi^\mathrm{spin}_{\rm max}\sim2.5$,
and $\chi^{p\mbox{-}\mathrm{orb}}_{d}(\bm 0)$ exceeds 
the spin susceptibility 
for $\chi^\mathrm{spin}_{\rm max}\gtrsim 10$ eV$^{-1}$.

To understand the origin of the 
enhancement of $\chi^{p\mbox{-}\mathrm{orb}}_d(\bm q)$,
we analyze the scaling flow of the effective interaction for the $p$O-CDW
introduced as
$\Gamma_d^{p\mbox{-}\mathrm{orb}}(\bm q)
\equiv 
\Gamma^c_{x;x}(\bm q)
+
\Gamma^c_{y;y}(\bm q)
-
\Gamma^c_{x;y}(\bm q)
-
\Gamma^c_{y;x}(\bm q)$ 
with 
$\Gamma^c_{\a;\b}(\bm q)
\equiv 
\sum_{\bm k,\bm k'}
\Gamma^c_{l}(\bm k+\bm q,\bm k; \bm k'+\bm q,\bm k')
\,
u^*_\a(\bm k+\q) \, u_\a(\bm k) \, \cdot
u_\b(\bm k'+\q) \, u_\b^*(\bm k') $.
Here $\Gamma^c_{l}$ is the charge-channel four-point vertex,
which is a moderate function of the Fermi momenta
in the parameter range of the present numerical study.
$u_\alpha(\bm k)$ is the  matrix element
connecting the $p$-orbitals ($\a =x,y$) and the conduction band
\cite{Tsuchiizu:2016ix}.
The scaling flow of 
$\Gamma_d^{p\mbox{-}\mathrm{orb}}(\bm q)$ is shown in 
Fig.\ \ref{fig3}(b), with the scaling parameter  
$l=\ln (\Lambda_0/\Lambda_l)$.
The negative effective interaction drives the enhancement of 
the corresponding instability.
We also plot the effective interaction for the spin channel,
$\Gamma^{\mathrm{spin}}(\bm Q_\mathrm{s})$.
For the spin-channel,
$\Gamma^{\mathrm{spin}}(\bm Q_\mathrm{s})\sim-U $ at $l=0$,
and it is renormalized like the RPA as
 $\Gamma^{\mathrm{spin}}_l=\Gamma^{\mathrm{spin}}_0/
(1-c |\Gamma^{\mathrm{spin}}_0|  l )$
for  $l\lesssim \ln (\Lambda_0/T)=1.6$, where $c$ is the density of states.
For the charge-channel,
although $\Gamma_d^{p\mbox{-}\mathrm{orb}}(\bm q)$ at $l=0$ is quite small,
it is strongly renormalized to be a large negative value.
This result means that the CDW instability originates from the VC
going beyond the RPA.

\begin{figure}[t]
\includegraphics[width=8cm]{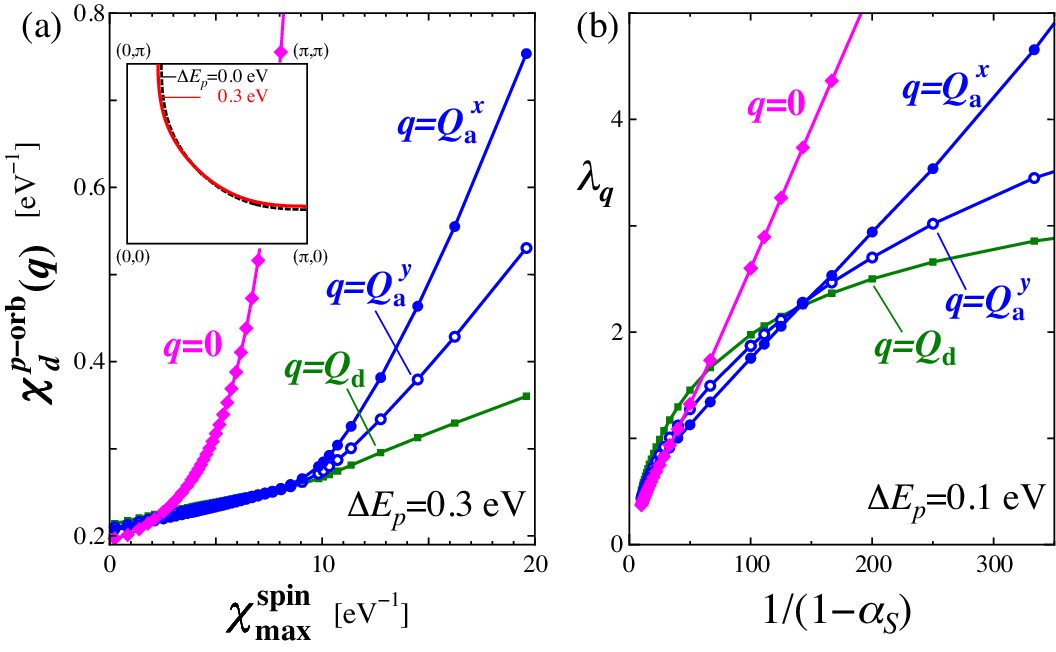}
\caption{
(a)
The RG+cRPA result of 
$\chi^{p\mbox{-}\mathrm{orb}}_d(\bm q)$ 
at $\bm q={\bm0}$, $\bm Q_\mathrm{a}^{x,y}$ and $\Q_\mathrm{d}$
as a function of $\chi^\mathrm{spin}_\mathrm{max}$ for $\Delta E_{p}=0.3$ eV.
The inset shows the FS.
(b) 
The eigenvalues of the CDW susceptibility
given by solving the linearized CDW equation
in Appendix D for $\gamma=0.3$ eV.
}
\label{fig4}
\end{figure}

We also calculate the 
$d$-electron charge susceptibility with form factor $f_\q(\bm k)$,
which is given as
\begin{equation}
\chi^{d\mbox{-}\mathrm{orb}} (\bm q)
=
\frac{1}{2}
\int_0^{1/T} d\tau \,
\left\langle 
B(\bm q,\tau)
B(-\bm q,0)
\right\rangle,
\label{eq:chid}
\end{equation}
where
$B (\bm q)= \sum_{\bm k,\sigma} f_\q(\bm k) \,
 d_{\bm k-\q/2,\sigma}^\dagger
 d_{\bm k+ \q/2, \sigma}$.
The numerically optimized $f_\q(\bm k)$ at $\q=\bm 0$ 
is shown in Fig.\ \ref{fig3}(c), which has the $B_{1g}$-symmetry.
Its  Fourier transformation 
gives the modulation of the effective hopping integrals,
called the $d_{x^2-y^2}$-wave bond-order.
Since $\k$-dependence of $f_{\bm0}(\bm k)$ 
in Fig.\ \ref{fig3}(c) is similar to that of $|u_x(\k)|^2-|u_y(\k)|^2$,
the obtained $\chi^\mathrm{spin}_\mathrm{max}$ dependence of 
$\chi^{d\mbox{-}\mathrm{orb}}({\bm 0})$
with the optimized form factor is similar to that of 
$\chi^{p\mbox{-}\mathrm{orb}}_{d}(\bm 0)$ shown in Fig.\ \ref{fig3}(a).
In the Appendix C, we analyze the single $d$-orbital Hubbard model,
and find the strong enhancement of $\chi^{d\mbox{-}\mathrm{orb}} (\bm q)$
with $B_{1g}$ form factor at $\q={\rm 0}$, $\Q_{\rm a}$ and $\Q_{\rm d}$,
very similarly to the $p$O-CDW susceptibilities shown in 
Figs. \ref{fig2} and \ref{fig3}.

As shown in Fig. \ref{fig2} (c), $\chi^{p\mbox{-}\mathrm{orb}}_d ({\bm 0})$  
increases divergently at $T\sim0.1$ eV, and the uniform 
$p$-orbital polarization with $n_{x}\ne n_{y}$
depicted in Fig.\ \ref{fig1}(c)
appears below the transition temperature.
In order to discuss the CDW instabilities \textit{inside} the 
nematic phase, we perform the RG+cRPA analysis 
in the presence of the uniform $p$O-CDW order
$H'=-\frac{1}{2}\Delta E_{p} [n_x(\bm 0)-n_y(\bm 0)]$.
In Fig.\ \ref{fig4}(a), we plot 
the peak values of $\chi^{p\mbox{-}\mathrm{orb}}_d(\bm q)$
in the uniform $p$O-CDW state with $\Delta E_{p}=0.3$ eV.
Due to small $\Delta E_{p}>0$,
$\chi^{p\mbox{-}\mathrm{orb}}_d(\bm q)$ at 
$\bm q =\bm Q_\mathrm{a}^x$ (along $x$-axis) strongly increases 
whereas that at $\bm q =\bm Q_\mathrm{a}^y$ (along $y$-axis) decreases.
Thus, the $p$O-CDW at $\bm q=\bm Q_\mathrm{a}^x$ is expected to emerge 
below $T_\mathrm{\rm CDW}$, consistently with the phase diagram in Fig.\ 
\ref{fig1}(a).

\section{Origin of Nematic Orders}

To confirm the mechanism of the nematic transition,
we perform the diagrammatic calculation for the MT- and AL-VCs.
These VCs can be obtained by solving the CDW equation 
introduced in Ref. \cite{Onari:2016gs}
in the study of Fe-based superconductors.
We analyze the linearized CDW equation
introduced in Appendix D
and in Ref. \cite{Kawaguchi-JPSJ}.
By solving the CDW equation,
both MT- and AL-VCs shown in Figs. \ref{fig2}(d) and \ref{fig2}(e)
with the optimized form factors
are systematically generated.
Figure \ref{fig4}(b) shows the eigenvalue of the linearized equation,
$\lambda_\q$, for $\Delta E_p=0.1$ eV.
Here, $\a_S$ is the spin Stoner factor, and 
the horizontal axis is proportional to $\chi^{\rm spin}_{\rm max}$.
The CDW susceptibility increases with the increase of $\lambda_\q$.
In Fig. \ref{fig4}(b), 
we set the quasiparticle damping $\gamma=0.3$ eV.
Note that $\lambda_\q$ decreases with $\gamma$,
whereas its overall $\q$-dependence is independent of $\gamma$,
as shown in Appendix D.
The obtained results are qualitatively consistent with 
the RG+cRPA results in Fig. \ref{fig4}(a).
In  Appendix D, we reveal that the CDW instabilities
at $\q={\bm 0}$ and $\q=\Q_{\rm a}$ are given by the 
higher-order AL-type VCs.

Finally, we explain why $\Lambda_{\rm low}$ should be set small.
The RG equation for the $\q={\bm0}$ vertex,
$\bar\Gamma^c_l(\k;\k')\equiv\Gamma^c_l(\k,\k;\k',\k')$, is given as
$d\bar\Gamma^c(\k;\k')/dl
\propto \Lambda_l \dot{f}(\Lambda_l)
\sum_{\k''}\delta(|E_{\k''}|-\Lambda_l)
\bar\Gamma^c_l(\k;\k'')\bar\Gamma^c_l(\k'';\k')$
\!+(other two terms),
where $\dot{f}(\e)$ is the derivative of the Fermi distribution function.
Since the factor $|\dot{f}(\Lambda_l)|$ is small for
$\Lambda_l \gtrsim 4T$,
the obtained $\bar\Gamma^c_l$ is strongly reduced
if $\Lambda_\mathrm{low}\gg T$.
In contrast, $\Gamma^{s,c}$ for $\q\ne\bm0$
is not so sensitive to $\Lambda_\mathrm{low}$.
For this reason, 
the renormalization effect of $\Gamma_l^c$ is underestimated
for $\q\approx\bm0$ if $\Lambda_{\rm low} \gg T$.
Then, the obtained $\chi^{p\mbox{-}\mathrm{orb}}_d(\bm 0)$ 
is suppressed to be
smaller than the peak values at $\q\ne\bm0$
if a large cutoff $\Lambda_\mathrm{low}\gg T$ is used,
similarly to the previous results 
for $\Lambda_\mathrm{low}=\pi T$ \cite{Tsuchiizu:2016ix}.
In the present study,
large $\chi^{p\mbox{-}\mathrm{orb}}_d(\bm 0)$ is correctly obtained 
thanks to the use of the small cutoff $\Lambda_\mathrm{low}=T$.

\section{Doping-dependence of the CDW susceptibilities}

In above sections, we studied the spin- and charge-susceptibilities
in the $d$-$p$ Hubbard model only for 10\% hole-doped case ($p=0.10$).
To understand the experimental phase diagram in Fig. \ref{fig1} (a),
however,
we have to study the doping dependence of susceptibilities.
This issue is a very important but difficult goal for theorists.
We find that the CDW is driven by the strong spin fluctuations, 
which strongly develop when the hole-density $p$ 
approaches zero experimentally.

\begin{figure}[t]
\includegraphics[width=0.65\linewidth]{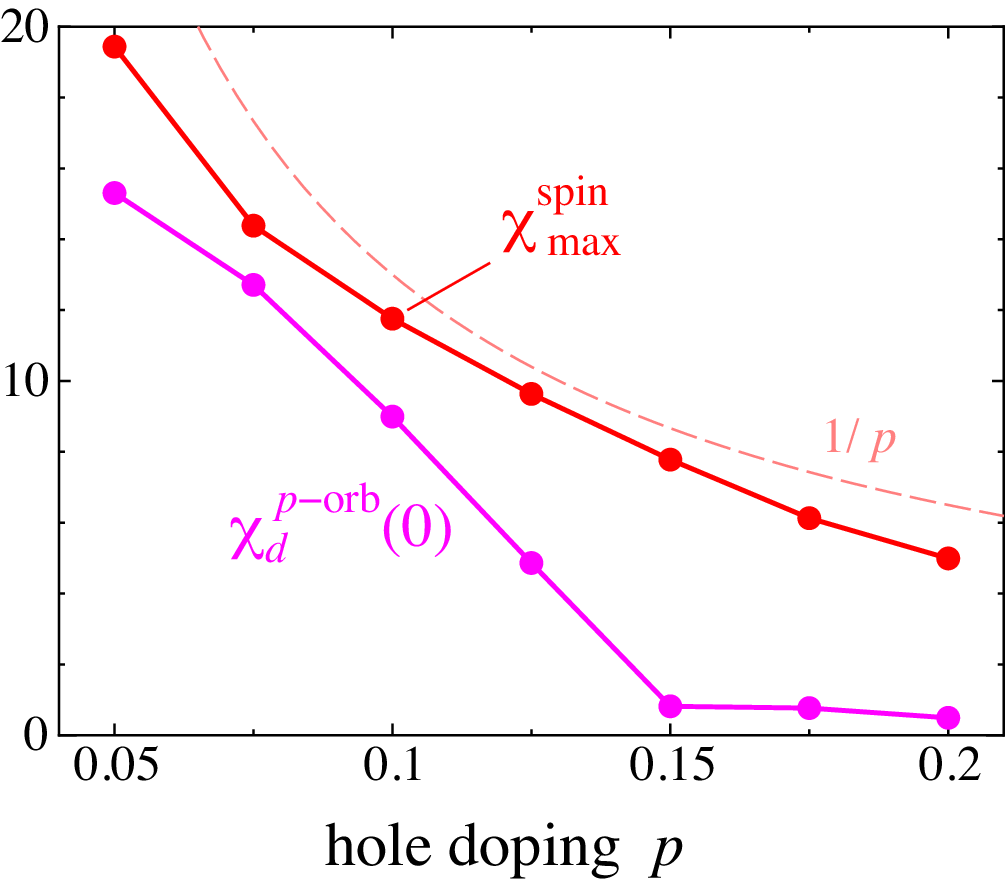}
\caption{
Doping dependence of $\chi^{p\mbox{-}\mathrm{orb}}_d(\bm 0)$
obtained by the RG+cRPA theory ($N_p=64$) at $T=0.1$ eV.
The obtained $\chi_d^{p\mbox{-}\mathrm{orb}}({\bm 0})$ linearly increases 
as $p$ approaches unity, consistently with $T^*$ 
in the experimental phase diagram in Fig. \ref{fig1} (a) in the main text. 
For $p\lesssim 0.1$, $\chi_d^{p\mbox{-}\mathrm{orb}}({\bm 0})$ exceeds 
$\chi_{\rm max}^{\rm spin}$ for $T\lesssim 0.1$ eV.
}
\label{fig5}
\end{figure}

In the RG+cRPA theory, the CDW is driven by the strong 
spin fluctuations, and spin fluctuations develop as the 
hole-carrier $p$ approaches zero experimentally.
For this reason, as shown in Fig. \ref{fig5},
the uniform CDW susceptibility $\chi_d^{p\mbox{-}\mathrm{orb}}(0)$ 
linearly increases as $p$ decreases
accompanied by the increment of $\chi_{\rm max}^{\rm spin}$ for $p\sim0$.
This result is consistent with the experimental 
$p$-dependence of $T^*$ in Fig. 1 (a) in the main text.
In Fig. \ref{fig5},
we modify $U$ slightly so that the experimental approximate relation 
$\chi_{\rm max}^{\rm spin} \propto 1/p$ is satisfied, 
as performed in our previous study \cite{Yamakawa:2015hb}.
We put $U=4.31$, $4.25$  and $4.06$ eV 
for $p=0.05$, $0.10$ and $0.20$, respectively.

In contrast to $T^*$, $T_{\rm CDW}$ decreases near the half-filling
for $p<0.1$ as depicted in Fig. \ref{fig1} (a). 
This behavior is also understood qualitatively based on the 
spin-fluctuation-driven mechanism. 
In fact, the axial CDW wavelength $\Q_{\rm a}$ is given by the 
nesting vector between the neighboring hot spots, 
and $|\Q_{\rm a}|$ increases as $p$ approaches zero.
The CDW instability driven by the Aslamazov-Larkin vertex correction,
which is qualitatively proportional to $\sum_\q\chi^s(\q)\chi^s(\q+\Q_{\rm a})$,
is suppressed if $|\Q_{\rm a}|$ is very large
as we explained in Ref. \cite{Yamakawa:2015hb}.
Therefore, the difference in the doping-dependences of
$T^*$ and $T_{\rm CDW}$ is qualitatively understood.
It is an important future issue to reproduce the 
experimental phase diagram in Fig. 1 (a) more completely,
which is one of the greatest goals in this field.

\section{Summary}

In summary, 
we studied various unconventional CDW instabilities 
in the $d$-$p$ Hubbard model by using the RG+cRPA method,
and predicted the multistage CDW transitions
in cuprate superconductors.
Based on the proposed spin-fluctuation-driven CDW mechanism,
the following understanding has been reached: 
The short-range spin-fluctuations drive the uniform nematic CDW 
around $T^*$, and it triggers the axial $\q=\Q_{\rm a}$ CDW 
at $T_{\rm CDW}$ successively.
We also explained the doping-dependence of $T^*$ 
based on the RG+cRPA theory.
These results naturally explain the phase diagram 
in Fig. \ref{fig1} (a), except for heavily under-doped region.
Although the uniform CDW order
cannot simply explain the pseudogap formation,
the large quasiparticle damping \cite{TPSC,Kotliar,Maier}
due to the short-range spin-fluctuations may induce the pseudogap  
for $T\lesssim T^*$.

\acknowledgements
We are grateful to Y. Matsuda, T. Hanaguri, T Shibauchi, Y. Kasahara,
Y. Gallais, W.\ Metzner, T. Enss, L. Classen, and S. Onari
for fruitful discussions.
This work was supported by Grant-in-Aid for Scientific Research from 
the Ministry of Education, Culture, Sports, Science, and Technology, Japan,
and in part by 
Nara Women's University Intramural Grant for Project Research.

\appendix

\section{RG equations for the four-point vertex}

In the main text,
we analyzed the $d$-$p$ Hubbard model by using the RG+cRPA method
\cite{Tsuchiizu:2015cs}.
This method is the combination of the fRG theory and the cRPA. 
The RG+cRPA method enables us to perform reliable numerical study 
in the unbiased way.
In this method, we introduce the original cutoff energy $\Lambda_{0}$
in order to divide each band 
into the higher and the lower energy regions:
The higher-energy scattering processes are calculated by using the cRPA:
The lower-energy scattering processes are analyzed by 
solving the RG equations,
in which the initial vertices in the differential equation
are given by the cRPA.

In the present model,
the bare Coulomb interaction term on $d$-electrons is given as
\begin{eqnarray}
&&\!\!\!\!\!\!\!\!\!\!\!\!
 H_{U}=\frac{1}{4}\sum_{i}
\sum_{\sigma \sigma' \rho \rho'}U^{0;\sigma \sigma' \rho \rho'}
d^{\dagger}_{i \sigma}d_{i \sigma'} d^{\dagger}_{i \rho'}d_{i \rho}  ,
\label{eqn:HUU} \\
&&\!\!\!\!\!\!\!\!\!\!\!\!
 U^{0;\sigma \sigma' \rho \rho'}
=\frac{1}{2}U^{0;s}
\vec{\bf{\sigma}}_{\sigma \sigma'} \cdot \vec{\bf{\sigma}}_{\rho' \rho}
+\frac{1}{2}U^{0;c}\delta_{\sigma,\sigma'}\delta_{\rho',\rho} ,
\label{eqn:HU}
\end{eqnarray}
where $U^{0;c}=U$ and $U^{0;s}=-U$.

\begin{figure}[b]
\includegraphics[width=.9\linewidth]{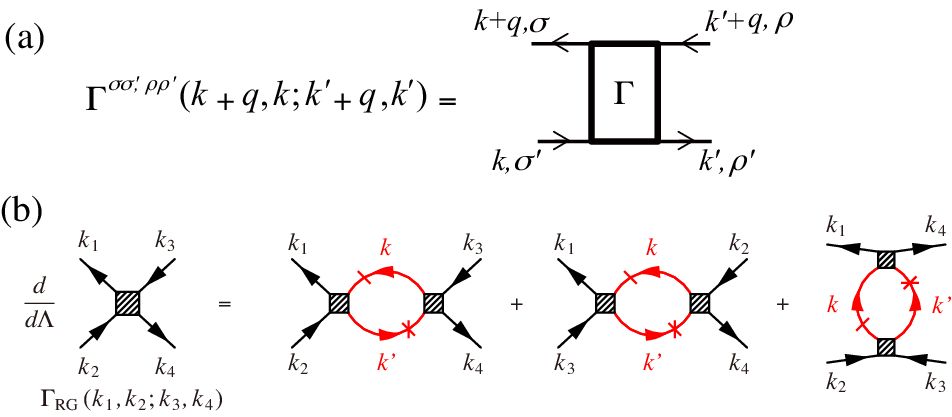}
\caption{
(a) Definition of the full four-point vertex 
$\Gamma^{\sigma \sigma' \rho \rho'}(\k+\q,\k;\k'+\q,\k')$ 
in the microscopic Fermi liquid theory.
(b) The one-loop RG equation for the four-point vertex.
The crossed lines represent the electron Green function with cutoff 
$\Lambda$.
The slashed lines represent the 
electron propagations having the energy shell $\Lambda$.
}
\label{fig:FS2}
\end{figure}

The antisymmetrized full four-point vertex
${\Gamma}(\k+\q,\k;\k'+\q,\k')$,
which is the dressed vertex of the
bare vertex ${\hat U}$ in Eq. (\ref{eqn:HU})
in the microscopic Fermi liquid theory,
is depicted in Fig. \ref{fig:FS2}(a).
Reflecting the SU(2) symmetry of the present model,
${\Gamma}$ is uniquely decomposed into the  
spin-channel and charge-channel four-point vertices
by using the following relation:
\begin{eqnarray}
&&\Gamma^{\sigma \sigma' \rho \rho'}(\k+\q,\k;\k'+\q,\k')
\nonumber \\
&& \ \ \ \ \ \ =\frac{1}{2}\Gamma^{s}(\k+\q,\k;\k'+\q,\k')
\vec{\bf{\sigma}}_{\sigma \sigma'} \cdot \vec{\bf{\sigma}}_{\rho' \rho}
\nonumber \\
&& \ \ \ \ \ \ \ 
+\frac{1}{2}\Gamma^{c}(\k+\q,\k;\k'+\q,\k')
\delta_{\sigma,\sigma'}\delta_{\rho',\rho} ,
\label{eqn:Gamma2}
\end{eqnarray}
where $\sigma, \sigma', \rho, \rho'$ are spin indices,
and $\vec{\bf{\sigma}}$ is the Pauli matrix vector.
We stress that ${\Gamma}^{c,s}$ 
are fully antisymmetrized, so the requirement by 
the Pauli principle is satisfied.
We note that 
${\Gamma}^{\uparrow\uparrow\uparrow\uparrow}
=\frac12 {\Gamma}^c+\frac12 {\Gamma}^s$,
${\Gamma}^{\uparrow\uparrow\downarrow\downarrow}
=\frac12 {\Gamma}^c-\frac12 {\Gamma}^s$,
and 
${\Gamma}^{\uparrow\downarrow\uparrow\downarrow}
={\Gamma}^s$.

In the RG formalism,
the four-point vertex function is determined 
 by solving the differential equations, called 
the RG equations.
In the band representation basis,
the explicit form of the RG equations is given by
\cite{Tazai:2016vd}:
\begin{eqnarray}
&& \frac{d}{d\Lambda}
\Gamma_{\rm RG}(k_1,k_2;k_3,k_4)
 = 
-\frac{T}{N}\sum_{k,k'}
\left[
\frac{d}{d\Lambda}
 G(k) \, G(k')
\right]
\nonumber \\
&& \ \ \ \ \ \ \times
\Bigl[
\Gamma_{\rm RG}(k_1,k_2;k,k') \, \Gamma_{\rm RG}(k,k';k_3,k_4)
\nonumber \\
&& \ \ \ \ \ \ {}
- \Gamma_{\rm RG}(k_1,k_3;k,k') \, \Gamma_{\rm RG}(k,k';k_2,k_4)
\nonumber \\
&& \ \ \ \ \ \
- \frac{1}{2} \Gamma_{\rm RG}(k_1,k; k',k_4) \, 
  \Gamma_{\rm RG}(k,k_2;k_3,k')
\Bigr] ,
\label{eqn:S-RG}
\end{eqnarray}
where $G(k)$ is the Green function multiplied by the 
Heaviside step function $\Theta(\Lambda-|E_{\k,\nu}|)$, and 
$k$ is the compact notation of the momentum, band, and spin index:
$k=(\k, \e_n, \nu, \sigma)$.
The diagrammatic representation of the RG equations 
is shown in Fig.\ \ref{fig:FS2}(b).
The first two contributions in the right-hand-side 
represent the particle-hole
  channels and the last contribution is the 
 particle-particle channel.

In a conventional fRG method,
$\Lambda_0$ is set larger than the bandwidth $W_{\rm band}$,
and the initial value is given by the bare Coulomb interaction 
in Eq. (\ref{eqn:HU}).
In the RG+cRPA method, 
we set $\Lambda_0<W_{\rm band}$, and the initial value is given by
the constrained RPA to include the higher-energy processes
without over-counting of diagrams \cite{Tsuchiizu:2015cs}.

In the main text,
we introduced the lower-energy cutoff $\Lambda_{\rm low}\ (\sim T)$
in the RG equation for the four-point vertex: Eq. (\ref{eqn:S-RG}).
For this purpose, 
we multiply the cutoff function $((\Lambda_{\rm low}/\Lambda)^\zeta+1)^{-1}$ 
to the right-hand-side of Eq. (\ref{eqn:S-RG}).
Here, $\zeta$ is a parameter determining the width 
of this smooth cutoff, and we set $\zeta=10$ in the main text.
We do not introduce the lower-energy cutoff 
in the RG equation for the susceptibilities.

\section{RG+cRPA analysis of the $d$-$p$ model with 
different hopping parameters}

\begin{figure}[t]
\includegraphics[width=\linewidth]{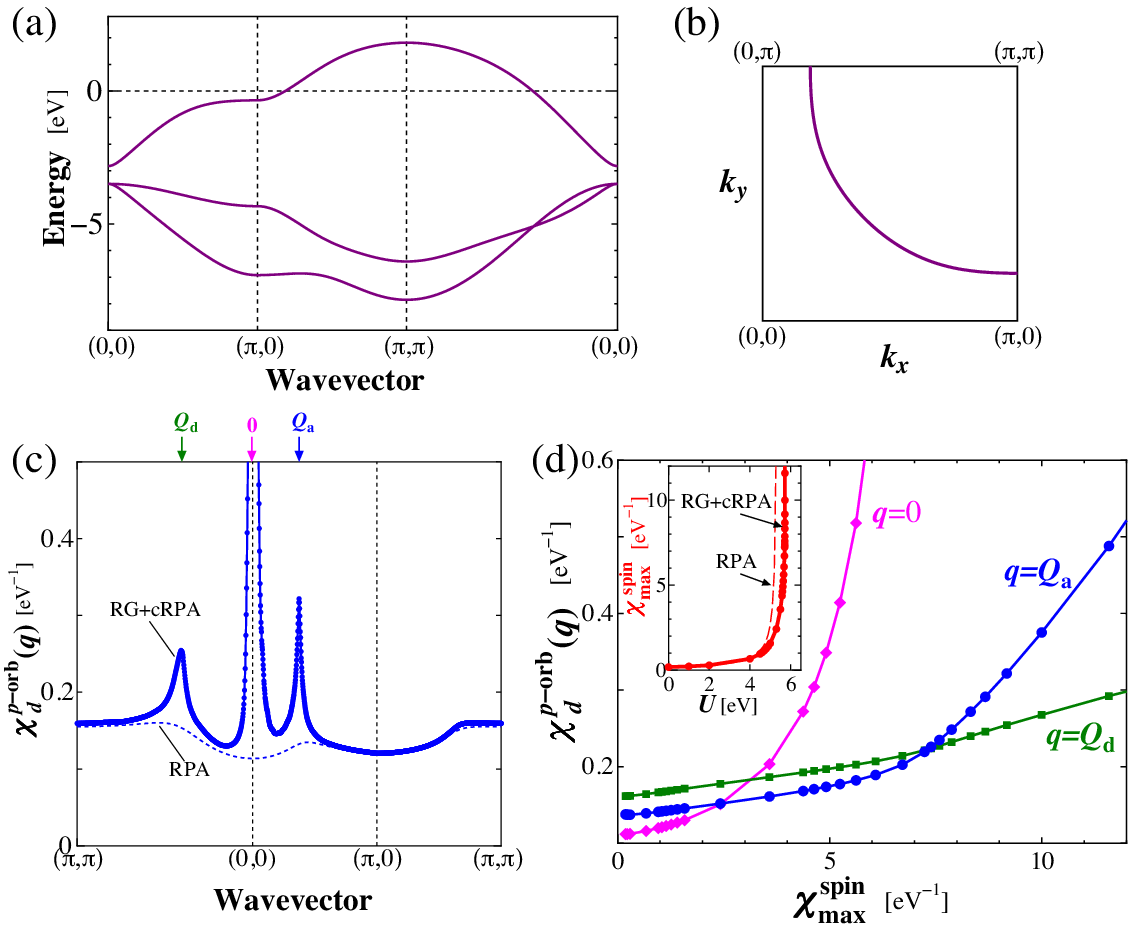}
\caption{
(a)  The energy dispersion and (b) the FS of the $d$-$p$ model 
with 1MTO model parameters.
(c) The RG+cRPA and RPA results for the $p$O-CDW susceptibility 
$\chi^{p\mbox{-}\mathrm{orb}}_d(\bm q)$
with $U=5.72$ eV.
(d) The RG+cRPA result of 
$\chi^{p\mbox{-}\mathrm{orb}}_d(\bm q)$ at three peak positions 
as a function of $\chi^\mathrm{spin}_\mathrm{max}$.
}
\label{fig:SM-N1MTO}
\end{figure}

In Ref. \cite{Hansmann:2014ib}, the authors derived 
the realistic $d$-$p$ models for La-based cuprate,
by evaluating the hopping parameters 
on the basis of the $N$th order muffin tin orbitals ($N$MTO).
The model parameters for $N=0$ and $N=1$ are given in 
Table \ref{tab:table}.
The bandstructure of the $N=0$ basis model (0MTO model) 
is very close to the LDA bandstructure near Fermi energy.
For this reason, we have analyzed the 0MTO model in the main text.
On the other hand, the $N=1$ basis model (1MTO model) appropriately
reproduces the overall oxygen bonding band structure 
with deep bottom energy $E \simeq -8$ eV.
In order to check the reliability of our RG+cRPA results, 
we analyze the $d$-$p$ model with the 1MTO model parameters.

Figure \ref{fig:SM-N1MTO}(a) shows the band structure of the 1MTO model.
Here, we introduced the third-nearest $d$-$d$ hopping 
$t_{dd}^\mathrm{3rd}=-0.1$ eV to make the FS closer to Y-based cuprates.
The FS of this model is shown in Fig. \ref{fig:SM-N1MTO}(b).
Now, we analyze this model by using the RG+cRPA method.
The parameters are the same as in the main text except for $U$.
The number of patches is $N_p=128$ and 
the initial cutoff is $\Lambda_0=0.5$ eV. 
The temperature is fixed at $T=0.1$ eV.

\begin{table}[t]
  \centering
  \begin{tabular}{ccccccccc}
\hline 
NMTO & $\e_d-\e_p$ & $t_{dd}$ & $t_{pd}$ & $t_{pd}'$ & $t_{pp}$ & $t_{pp}'$ & $t_{pp}''$ & $t_{pp}'''$ \\
\hline \hline
$N=0$ & $0.43$ & $-0.10$ & $0.96$ & $-0.10$ & $0.15$ & $-0.24$ & $0.02$ & $ 0.11$ \\
\hline 
$N=1$ & $0.95$ & $0.15$ & $1.48$ & $0.08$ & $0.91$ & $0.03$ & $0.15$ & $0.03$ \\
\hline 
  \end{tabular}
  \caption{
Hopping integrals for the $N=0$ and $N=1$ models 
given in Ref. \cite{Hansmann:2014ib}.
The unit is eV.
}
  \label{tab:table}
\end{table}

In Fig. \ref{fig:SM-N1MTO}(c), 
we show the obtained $\chi^{p\mbox{-}\mathrm{orb}}_{d}(\bm q)$
for $U=5.72$ eV.
The RPA results are also shown for comparison.
It has the largest peak  at $\bm q=\bm 0$ 
and the second largest peak at $\bm q=\bm Q_\mathrm{a}$,
respectively.
The obtained $\bm q$-dependence of 
$\chi^{p\mbox{-}\mathrm{orb}}_{d}(\bm q)$ is similar to Fig.\ 2(b).
We also investigate the  $U$-dependences of
the spin and charge susceptibilities.
As shown in the inset of Fig.\ \ref{fig:SM-N1MTO}(d),
relatively large $U$ is required for the enhancement of 
$\chi^\mathrm{spin}_{\rm max}$ in the 1MTO model,
since the density of states of the $d$ orbital at the Fermi energy 
in the 1MTO model is  smaller than  that in the 0MTO model
\cite{Hansmann:2014ib}.
In Fig.\ \ref{fig:SM-N1MTO}(d),
we plot the peak values of 
$\chi^{p\mbox{-}\mathrm{orb}}_d(\bm q)$ as functions of 
$\chi^\mathrm{spin}_{\rm max}$.
The obtained results are quite similar to Fig.\ 3(a) in the main text. 
Thus the spin-fluctuation-driven CDW instabilities 
 are universal phenomena in both 
 the 0MTO model and 1MTO model.

In summary, we investigate the $d$-$p$ model with 
1MTO model parameters.
We found that the results are 
very similar to those for the 0MTO model given in the main text.
Therefore the 
mechanism of the spin-fluctuation-driven CDW instabilities 
revealed in the main text is universal,
independently of the details of the model parameters.

\section{RG+cRPA analysis for the single $d$-orbital Hubbard model}

In the main text, 
we studied the 0MTO $d$-$p$ Hubbard model based on the RG+cRPA theory,
and found that the $p$O-CDW susceptibilities develop strongly
in the strong spin-fluctuation region.
Similar results are obtained in the 1MTO model
in which $\e_d-\e_p$ is 0.53eV larger than that in the 0MTO model,
as we show in Appendix B.
In these $d$-$p$ models, any Coulomb interactions on $p$-orbitals
are not taken into account.
Therefore, spin-fluctuation-driven CDW formation 
is also expected to be realized in the single $d$-orbital 
Hubbard model with on-site Coulomb interaction.

\begin{figure}[t]
\includegraphics[width=0.99\linewidth]{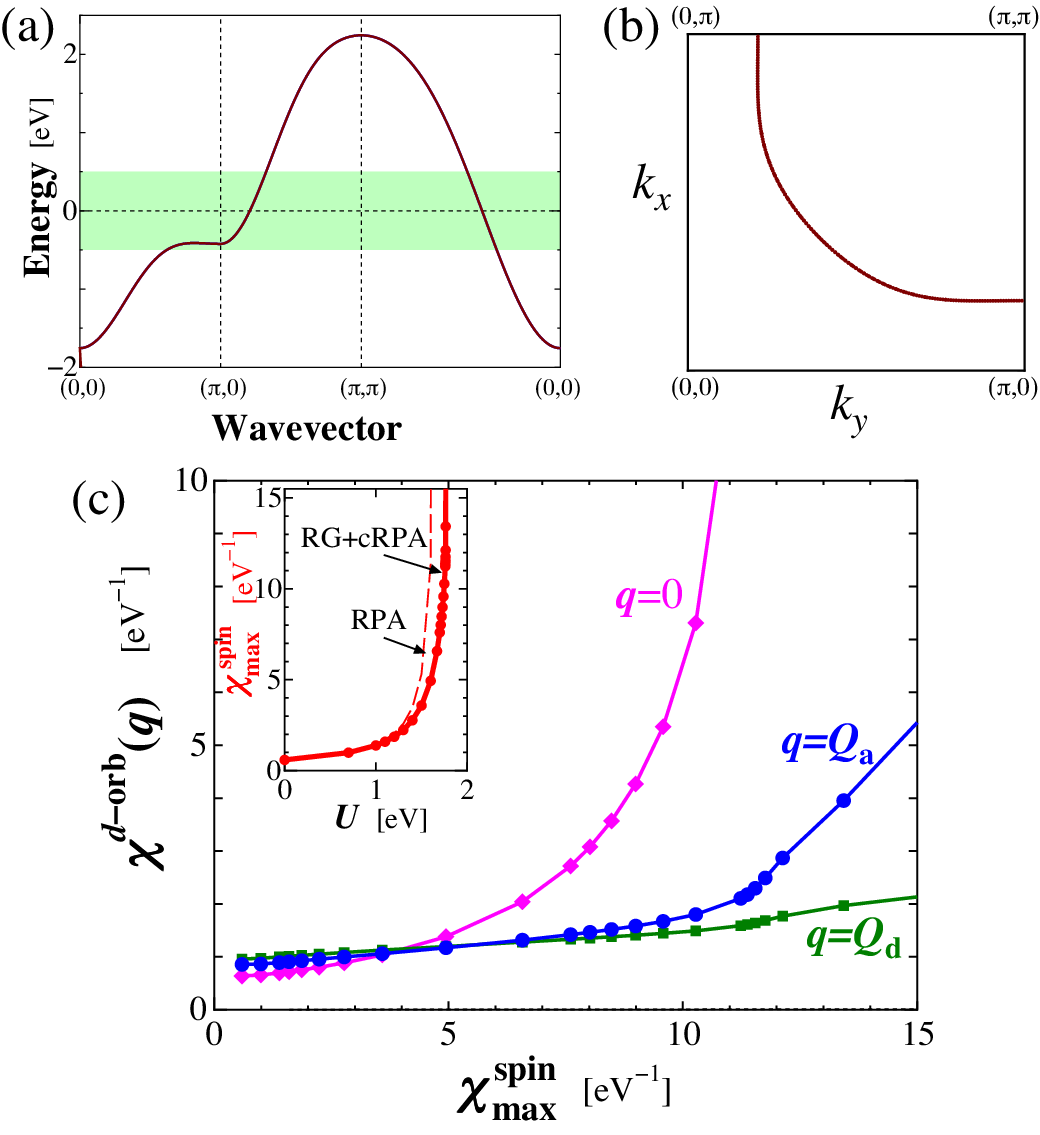}
\caption{
(a) The energy dispersion and (b) the FS of the 
single $d$-orbital Hubbard model.
(c) The RG+cRPA result of the $d$-electron charge susceptibility
with $B_{1g}$ form factor $f_\q(\k)=\cos(k_x)-\cos(k_y)$,
$\chi^{d\mbox{-}\mathrm{orb}}_d(\bm q)$,
as a function of $\chi^\mathrm{spin}_\mathrm{max}$.
}
\label{fig:SM-Hub}
\end{figure}

Here, we study the single $d$-orbital Hubbard model
with the first-, the second-, and the third-nearest hopping integrals
as $t=-0.50$ eV, $t'=0.083$ eV, and $t'''=-0.10$ eV, respectively. 
The bandstructure and Fermi surface for $n=0.90$ are shown in
Figs. \ref{fig:SM-Hub} (a) and (b), respectively.
We calculate the $d$-electron charge susceptibility 
$\chi^{d\mbox{-}\mathrm{orb}}(\bm q)$ 
with the $B_{1g}$ form factor $f_{\bm q}(\bm k)=\cos(k_x) - \cos(k_y)$
introduced in Eq.\ (3) in the main text,
The obtained results are summarized in Fig. \ref{fig:SM-Hub} (c):
Both $\chi^{d\mbox{-}\mathrm{orb}}({\bm 0})$
and $\chi^{d\mbox{-}\mathrm{orb}}(\Q_{\rm a})$ are 
strongly enlarged in the strong spin-fluctuation region,
very similarly to the $p$O-CDW susceptibility
shown in Fig. 3 (a) in the main text.

Therefore, it was verified that our main numerical results in the main text 
are unchanged even in the single $d$-orbital model, 
once the $B_{1g}$ form factor is taken into account. 
We also analyzed the CDW equation for the single $d$-orbital model,
and obtained the strong CDW instability.
The obtained form factor $\Delta\Sigma_{{\bm 0}}(\k)$ has the $B_{1g}$-symmetry.
In real space, this is the bond-order (=modulation of the hopping integrals)
given by the Fourier transformation of the 
symmetry-breaking self-energy $\Delta\Sigma_{{\bm 0}}(\k)$.
Thus, the robustness of the spin-fluctuation-driven CDW mechanism has 
been clearly confirmed.

\section{Analysis of the linearized CDW equation}

In the main text, we analyzed the $d$-$p$ Hubbard model 
for cuprate superconductors in an unbiased manner
using the RG+cRPA method.
We find that the  nematic CDW with $d$-form factor 
is the leading instability. 
The axial nematic CDW instability at $\q=\Q_{\rm a}$ 
is the second strongest, and its strength increases under the 
static uniform CDW order.
This result leads to the prediction that uniform nematic CDW occurs at 
the pseudogap temperature $T^*$, and the axial CDW at wavevector 
$\q=\Q_{\rm a}$ is induced under $T^*$.

\begin{figure}[t]
\includegraphics[width=.9\linewidth]{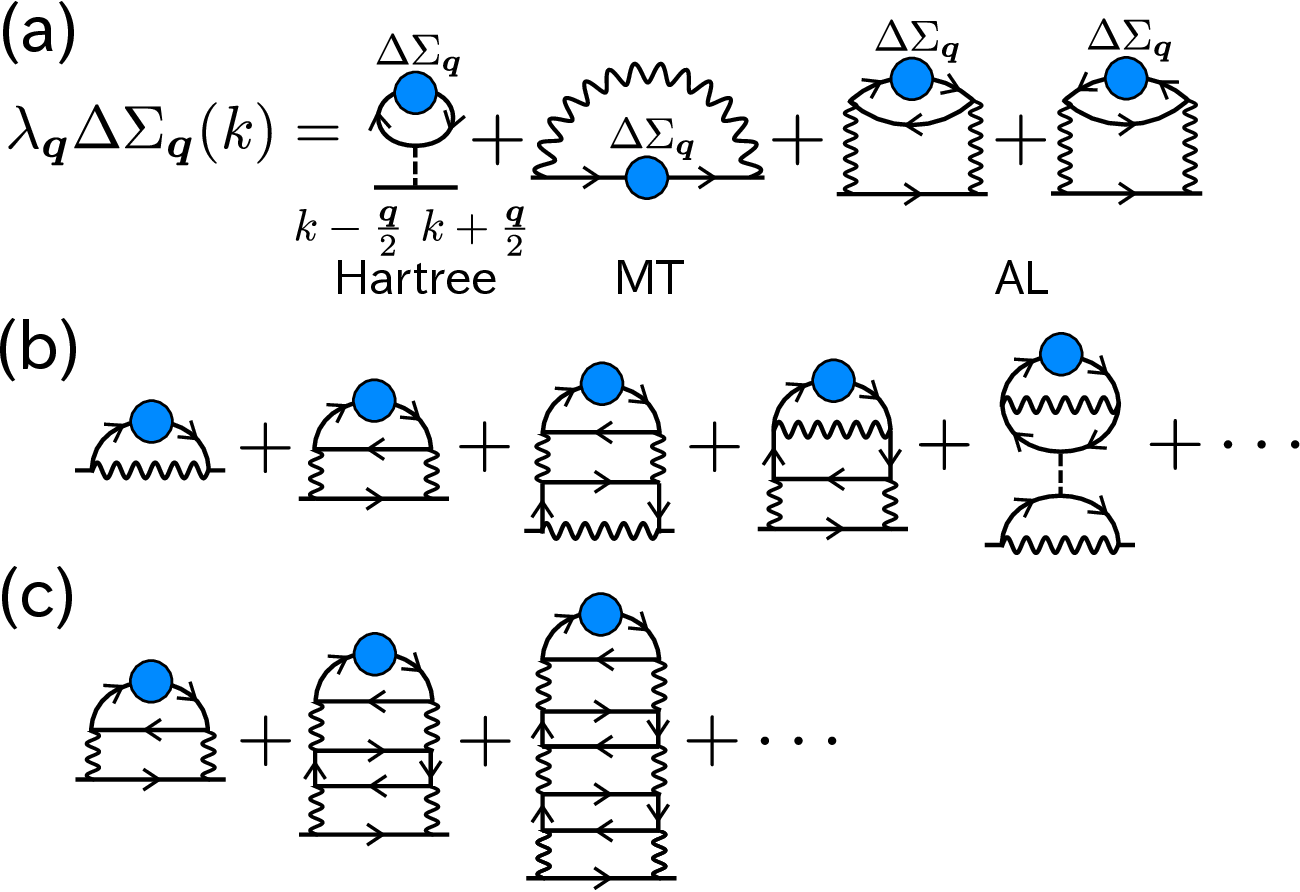}
\caption{
(a) Schematic linearized CDW equation for general wavevector $\q$.
(b) Examples of the VCs generated by solving the linearized CDW equation.
(c) Higher-order AL processes.
}
\label{fig:figS1}
\end{figure}

In this section,
we study the CDW formation mechanism in cuprate superconductors
based on the diagrammatic method,
in order to find what many-body processes cause the CDW order.
Theoretically, the CDW order is given as the
symmetry-breaking in the self-energy $\Delta\Sigma(k)$.
According to Refs. \cite{Onari:2012jb,Kawaguchi-JPSJ},
the self-consistent CDW equation is given as
\begin{eqnarray}
\Delta\Sigma(k)= (1-P_{\rm A_{1g}})T\sum_q V(q)G(k+q),
\label{eqn:self}
\end{eqnarray}
where 
$P_{\rm A_{1g}}$ is the $A_{\rm 1g}$-symmetry projection operator, and
$G(k)= (G_0^{-1}(k)-\Delta \Sigma(k))^{-1}$ is the $d$-electron Green function 
with the symmetry-breaking term $\Delta \Sigma$.
$V(q)=U^2(\frac32 \chi^s(q)+\frac12 \chi^c(q)-\chi^0(q))+U$, 
where 
${\chi}^{s(c)}(q)={\chi}^{0}(q)/({1}-(+){U}{\chi}^{0}(q))$ and
$\chi^{0}(q)=-T\sum_{k}G(k+q)G(k)$.

In order to analyze the CDW state with arbitrary wavevector $\q$,
we linearize Eq. (\ref{eqn:self}) with respect to $\Delta\Sigma$:
\begin{eqnarray}
&& \lambda_\q \Delta\Sigma_\q(k)= T\sum_{k'} K(\q;k,k')\Delta\Sigma_\q(k'),
\label{eqn:linearized}    
\end{eqnarray}
where $\lambda_\q$ is the eigenvalue for the CDW for the wavevector $\q$.
The CDW with wavevector $\q$ appears when $\lambda_\q=1$, and
the eigenvector $\Delta\Sigma_\q(k)$ gives the CDW form factor.
The kernel $K(\q,k,k')$ is given in Fig. \ref{fig:figS1}(a),
and its analytic expression is given as \cite{Kawaguchi-JPSJ}:
\begin{eqnarray}
&& \!\!\!\!\!\!\!\!\!\!\!
K(\q;k,k')=
\left(\frac32 V_0^s(k-k')+\frac12 V_0^c(k-k')\right)
\nonumber \\
&& \ \ \ \ \ \ \ \ \ \ 
\times G_0(k'+{\q}/{2})G_0(k'-{\q}/{2})
 \nonumber \\
&&  \ \ \ \
-T\sum_p \left( \frac32 V_0^s(p+{\q}/{2})V_0^s(p-{\q}/{2}) \right.
 \nonumber \\
&& \ \ \ \ \ \ \ \
\left. +\frac12 V_0^c(p+{\q}/{2})V_0^c(p-{\q}/{2}) \right)
\nonumber \\
&& \ \ \ \ \ \ \ \ \ \ 
\times G_0(k-p) \left(\Lambda_\q(k';p)+\Lambda_\q(k';-p)\right) ,
\label{eqn:S-K} 
\end{eqnarray}
where 
$V_0^s(q)=U+U^2\chi_0^s(q)$, $V_0^c(q)=-U+U^2\chi_0^c(q)$, and
$\Lambda_\q(k;p)\equiv G_0(k+\frac{\q}{2})G_0(k-\frac{\q}{2})G_0(k-p)$.
The subscript $0$ in Eq. (\ref{eqn:S-K}) represents the 
functions with $\Delta\Sigma=0$.

By solving the linearized CDW equation (\ref{eqn:linearized}),
many higher-order vertex corrections (VCs) 
are systematically generated.
Some examples of the generated VCs are shown in Fig. \ref{fig:figS1}(b).
If we drop the Hartree term and MT term in $K(\q;k,k')$,
we obtain the series of higher-order AL-VCs shown in Fig. \ref{fig:figS1}(c).
The AL terms drive the $\q={\bm0}$ CDW instability 
since its functional form $\propto \sum_\k\chi^s(\k+\q)\chi^s(\k)$ 
is large for $\q\approx{\bm0}$
\cite{Onari:2016gs}.

\begin{figure}[t]
\includegraphics[width=.99\linewidth]{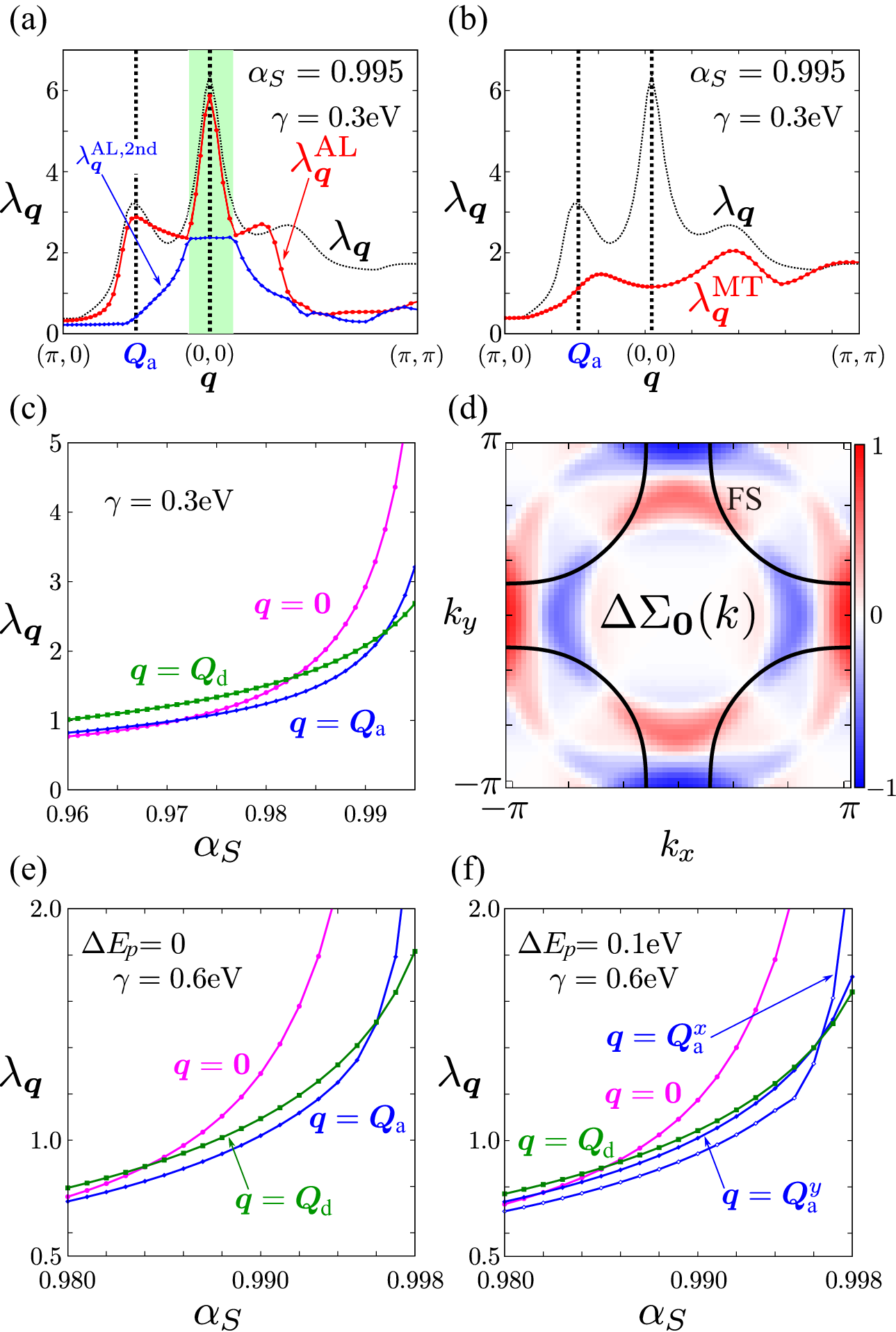}
\caption{
(a)(b) $\q$-dependences of 
$\lambda_\q$, $\lambda_\q^{\rm AL}$, $\lambda_\q^{\rm AL,2nd}$, 
and $\lambda_\q^{\rm MT}$ for $\alpha_S=0.995$ and $\gamma=0.3$ eV.
(c) $\lambda_\q$ at $\q=$ $\bm0$, $\Q_{\rm a}$ and $\Q_{\rm d}$ 
as function of $\a_S$ for $\gamma=0.3$ eV.
(d) Form factor for $\q={\bm 0}$ ($d$-wave).
(e)(f) $\lambda_\q$ as function of $\a_S$ for $\gamma=0.6$ eV
in the cases of $\Delta E_p=0$ and $\Delta E_p=0.1$ eV, respectively.
}
\label{fig:figS2}
\end{figure}

Figure \ref{fig:figS2}(a) shows the obtained 
$\q$-dependence of $\lambda_\q$ for $\a_S=0.995$ at $T=50$ meV.
Here, we introduced the quasiparticle damping $\gamma=0.3$ eV into $G_0(k)$.
Here, $\lambda_\q$ is the largest at $\q={\bm 0}$, and
the second largest maximum is at $\q=\Q_{\rm a}$.
$\a_S\equiv U{\max}_\q \{\chi^0_0(\q)\}$ is the spin Stoner factor.
We also show the eigenvalue $\lambda^{\rm AL}_\q$
(and the second-largest eigenvalue $\lambda^{\rm AL, 2nd}_\q$),
which is obtained by dropping the Hartree and MT terms in the kernel.
That is, $\lambda^{\rm AL}_\q$ is given by the 
higher-order AL processes shown in Fig. \ref{fig:figS1}(c).
At $\q={\bm0}$ and $\Q_{\rm a}$,
$\lambda^{\rm AL}_\q$ is almost equal to 
the true eigenvalue $\lambda_\q$.

In the present analysis,
we dropped the $\e_n$-dependence of $\Delta\Sigma_\q(k)$
by performing the analytic continuation $(i\e_n \rightarrow \e)$
and putting $\e=0$.
We also dropped the $\e_n$-dependence of the 
quasiparticle damping $\gamma$.
Due to these simplifications, 
the obtained $\lambda_\q$ is overestimated.
Therefore, we do not put the constraint $\lambda_\q<1$ here.


In Fig. \ref{fig:figS2}(b), we show the eigenvalue $\lambda^{\rm MT}_\q$,
which is obtained by dropping the Hartree and AL terms in the kernel.
It is much smaller than $\lambda_\q$ at $\q={\bm0}$ and $\Q_{\rm a}$,
whereas  $\lambda_\q$ at $\q=\Q_{\rm d}$ is comparable to the true eigenvalue.
Therefore, the origin of the CDW instability at $\q={\bm0}$ and $\Q_{\rm a}$
is the AL process, whereas that at $\q=\Q_{\rm d}$ is mainly the MT process.

Figure \ref{fig:figS2} (c) shows the eigenvalues 
at $\q={\bm 0}$, $\Q_{\rm a}$, and $\Q_{\rm d}$ as function of $\a_S$.
As the spin susceptibility increases ($\a_S\gtrsim0.98$),
$\lambda_\q$ is drastically enlarged by the VCs,
and $\lambda_{\q={\bm0}}$ becomes the largest due to the AL processes.
The form factor at $\q={\bm0}$, $\Delta\Sigma_{{\bm 0}}(\k)$,
has the $d$-wave symmetry as shown in Fig. \ref{fig:figS2} (d).

We stress that the eigenvalue $\lambda_\q$ is quickly suppressed
by increasing $\gamma$, which is actually large in cuprates. 
Figures \ref{fig:figS2} (e) and (f) show the 
CDW susceptibilities for larger damping rate $\gamma=0.6$ eV,
in the cases of $\Delta E_p=0$ and $\Delta E_p=0.1$ eV, respectively.
 (Note that the damping rate is renormalized to be $\sim\gamma/5$ in cuprates.)
In Fig. \ref{fig:figS2} (e), $\lambda_\q$ reaches unity first at $\q={\bm 0}$ 
with increasing $\alpha_S$. 
In the nematic state with $\Delta E_p=0.1$ eV shown in Fig. \ref{fig:figS2} (f),
$\lambda_\q$ at $\q=\Q_{\rm a}^x$ exceeds $\lambda_{\Q_{\rm d}}$ for $\a_S>0.996$.
The corresponding eigenvalue is $\sim1.4$,
which decreases with increasing $\gamma$.
This result supports the main result of the present RG+cRPA study 
shown in Fig. 4 (a) in the main text.

In summary,
we analyzed the linearized CDW equation
based on the $d$-$p$ Hubbard model, 
by including both the MT and AL VCs into the kernel.
When the spin fluctuations are strong ($\a_S\gtrsim0.98$),
the uniform nematic CDW has the strongest instability.
The axial CDW instability is strongly magnified 
under the uniform CDW order,
as we explain the main text.
The obtained results are consistent with the results by the RG+cRPA
in the main text.
Thus, it is concluded that the higher-order AL processes
give the CDW orders at $\q={\bm0}$ and $\Q_{\rm a}$.



\end{document}